\documentclass[twocolumn,showpacs,preprintnumbers,
               superscriptaddress,floatfix]{revtex4}
\usepackage{graphicx}

\newcommand{\beq}{\begin{eqnarray}}
\newcommand{\eeq}{\end{eqnarray}}

\newcommand{\la}{\langle}
\newcommand{\ra}{\rangle}
\newcommand{\qc}{\la \bar{q}q \ra}
\newcommand{\uc}{\la \bar{u}u \ra}
\newcommand{\dc}{\la \bar{d}d \ra}
\newcommand{\s}{\la \bar{s}s \ra}

\newcommand{\dmdmu}{{d m \over d \mu}}

\newcommand{\dqcdmu}{{d \qc \over d \mu}}
\newcommand{\ducdmu}{{d \uc \over d \mu}}
\newcommand{\ddcdmu}{{d \dc \over d \mu}}
\newcommand{\dscdmu}{{d \s \over d \mu}}
\newcommand{\dqcdmus}{{d \qc \over d \mu_S}}

\newcommand{\dqcdmuv}{{d \qc \over d \mu_V}}

\newcommand{\ddqcdmu}{{d^2 \qc \over d \mu^2}}
\newcommand{\ddqcdmus}{{d^2 \qc \over d \mu_S^2}}
\newcommand{\ddqcdmuv}{{d^2 \qc \over d \mu_V^2}}

\begin{document}
%\draft
%
%------------------------- Title ------------------------------------%
%--------------------------------------------------------------------%
%\preprint{\vbox{ \hfill }}
\title{Responses of quark condensates to the chemical potential}

\author{O. Miyamura\footnote{Deceased.}}
\affiliation{Dept. of Physics, Hiroshima University,
            Higashi-Hiroshima 739-8526, Japan}

\author{S. Choe}
\affiliation{Dept. of Chemistry, Korea Advanced Institute of
              Science and Technology, Daejon 305-701, Korea}

\author{Y. Liu}
\affiliation{Dept. of Physics, Hiroshima University,
            Higashi-Hiroshima 739-8526, Japan}

\author{T. Takaishi}
\affiliation{Hiroshima University of Economics,
              Hiroshima 731-01, Japan}

\author{A. Nakamura}
\affiliation{IMC, Hiroshima University,
             Higashi-Hiroshima 739-8521, Japan}

%\date{}

\begin{abstract}
The responses of quark condensates to the chemical potential, as a
function of temperature $T$ and chemical potential $\mu$, are
calculated within the Nambu--Jona-Lasinio (NJL) model. We compare
our results with those from the recent lattice QCD simulations
[QCD-TARO Collaboration, Nucl. Phys. B (Proc. Suppl.)
\textbf{106}, 462 (2002)]. The NJL model and lattice calculations
show qualitatively similar behavior, and they will be
complimentary ways to study hadrons at finite density. The
behavior above $T_c$ requires more elaborated analyses.
\end{abstract}


\maketitle

%------------------------ Text --------------------------------------%
%--------------------------------------------------------------------%
The variation of quark condensates in medium plays a key role to
understand the behavior of hadron masses and chiral symmetry
restoration \cite{br91,hl92hkl93}. Recently, we calculated for the
first time the second order response of the quark condensate to
the chemical potential $\ddqcdmu$ at $\mu$ = 0 using lattice QCD
\cite{taro-lattice} following the method in \cite{taro-prd}. It
was found that the response is negative both below and above
$T_c$. One of interesting results is that the response to the
isoscalar chemical potential ($\mu_S$ = $\mu_u$ = $\mu_d$) is
almost the same as that to the isovector chemical potential
($\mu_V$ = $\mu_u$ = -- $\mu_d$), where $\mu_u$ ($\mu_d$) is the
$u$ ($d$) quark chemical potential. It would be interesting if we
can check this result within effective models of QCD. In this work
we present an Nambu--Jona-Lasinio (NJL) model \cite{njl,vw91k92}
calculation of $\dqcdmu$ and $\ddqcdmu$ at $\mu$ = 0 and compare
our results with those from the lattice QCD simulations
\cite{taro-lattice}.

An SU(2) NJL model Lagrangian will be enough for that purpose.
However, we are also interested in the responses of the s-quark
condensate and a comparison with those of the u,d-quark
condensates will be useful for future studies in the lattice
calculations. Thus we used an SU(3) NJL model in our calculations
and found that the effects of the flavor mixing are negligible. We
present only the results for the u,d-quark condensates in this
paper, and we will make the comparison in a forthcoming paper.

First, let us consider an SU(3) NJL model Lagrangian \cite{hk94}:
\beq
 L &=& \bar{q}(i\gamma \cdot d - m)q
 + {1 \over 2} g_S \sum_{a=0}^8 \left[ (\bar{q}\lambda_aq)^2 +
(\bar{q} i\lambda_a\gamma_5q)^2 \right]
 \nonumber \\
 &+& g_D \left[ {\rm det} ~\bar{q}_i (1 - \gamma_5)q_j + h.c. \right]  ,
 \label{lag}
\eeq
where $\lambda_a$ are the Gell-Mann matrices and $m$ is a mass
matrix for current quarks, $m$=diag($m_u$, $m_d$, $m_s$). We take
the following parameters in \cite{hk94}:
\beq \Lambda = 631.4 ~{\rm MeV}, ~g_S\Lambda^2 = 3.67,
~g_D\Lambda^5 = -9.29
\nonumber \\
m_u = m_d = 5.5 ~{\rm MeV}, ~m_s = 135.7 ~{\rm MeV}  ,
\label{para1}
 \eeq
where $\Lambda$ is the momentum cut-off. The third term in
Eq.(\ref{lag}) is a reflection of the axial anomaly, and causes a
mixing in flavors. For example, the constituent quark masses are
given as follows.
\beq
M_u &=& m_u - 2 g_S \uc - 2 g_D \dc\s , \nonumber \\
M_d &=& m_d - 2 g_S \dc - 2 g_D \uc\s , \nonumber \\
M_s &=& m_s - 2 g_S \s  - 2 g_D \uc\dc ,
 \label{mass} \eeq
where $\la \cdot \ra$ means the statistical average. In this work
we concentrate mostly on the Case II in \cite{hk94}, where only
$g_D$ has a temperature dependence
\beq g_D(T) = g_D(T=0) ~{\rm exp} [-(T/T_0)^2] ,
 \label{gd}
 \eeq
while other coupling constants and the cut-off are independent of
$T$ and chemical potential (or density). Here, we set $T_0$ = 0.1
GeV taking into account the restoration of $U_A(1)$ symmetry as in
\cite{hk94}.

In the mean-field approximation the above Lagrangian leads to the
following gap equation \cite{hk94}:
\beq
 Q_i  \equiv 2 N_c \sum_p \left( {-M_i \over E_{ip}} f(E_{ip}) \right)
 = \la \bar{q}_iq_i \ra ,
 \label{gap}
\eeq
where the index $i$ denotes the $u$, $d$, and $s$ quarks. $N_c$ is
the number of colors and $M_i$ is the constituent quark mass, and
$E_{ip}=\sqrt{M_i^2 + p^2}$. $f(E_{ip})=1-n_{ip}-\bar{n}_{ip}$,
where $n_{ip}$ and $\bar{n}_{ip}$ are the distribution functions
of the $i$th quark and antiquark, respectively, i.e.,
\beq n_{ip} = {1  \over 1 + {\rm exp}~((E_{ip} - \mu_i)/T)} ,
 \nonumber \\
 \bar{n}_{ip} = {1 \over 1 + {\rm exp}~((E_{ip} + \mu_i)/T)} .
\eeq

The l.h.s. of Eq.(\ref{gap}) is a function of $\uc$, $\dc$, $\s$,
$\mu_i$, and $T$. Then, we obtain the first order response of
quark condensates $\ducdmu$, $\ddcdmu$, and $\dscdmu$ by
differentiating both sides with respect to $\mu$ at a fixed $T$,
i.e., we solve the following equations
\beq
 {d Q_u (\uc, \dc, \s, \mu_u) \over
d \mu} &=& {d \uc \over d \mu} ,
 \nonumber \\
{d Q_d (\uc, \dc, \s, \mu_d) \over d \mu} &=& {d \dc \over d \mu}
,
\nonumber \\
{d Q_s (\uc, \dc, \s, \mu_s) \over d \mu} &=& {d \s \over d \mu} .
 \label{dqdmu}
 \eeq
Here, we consider two types of $\mu$ following the notation in the
lattice QCD simulations \cite{taro-prd,taro-lattice}. One is the
isoscalar chemical potential $\mu_S$ = $\mu_u$ + $\mu_d$, and the
other is the isovector one $\mu_V$ = $\mu_u$ -- $\mu_d$.
%------------------
\begin{figure*}[!]
  \includegraphics[height=8cm,width=8cm]{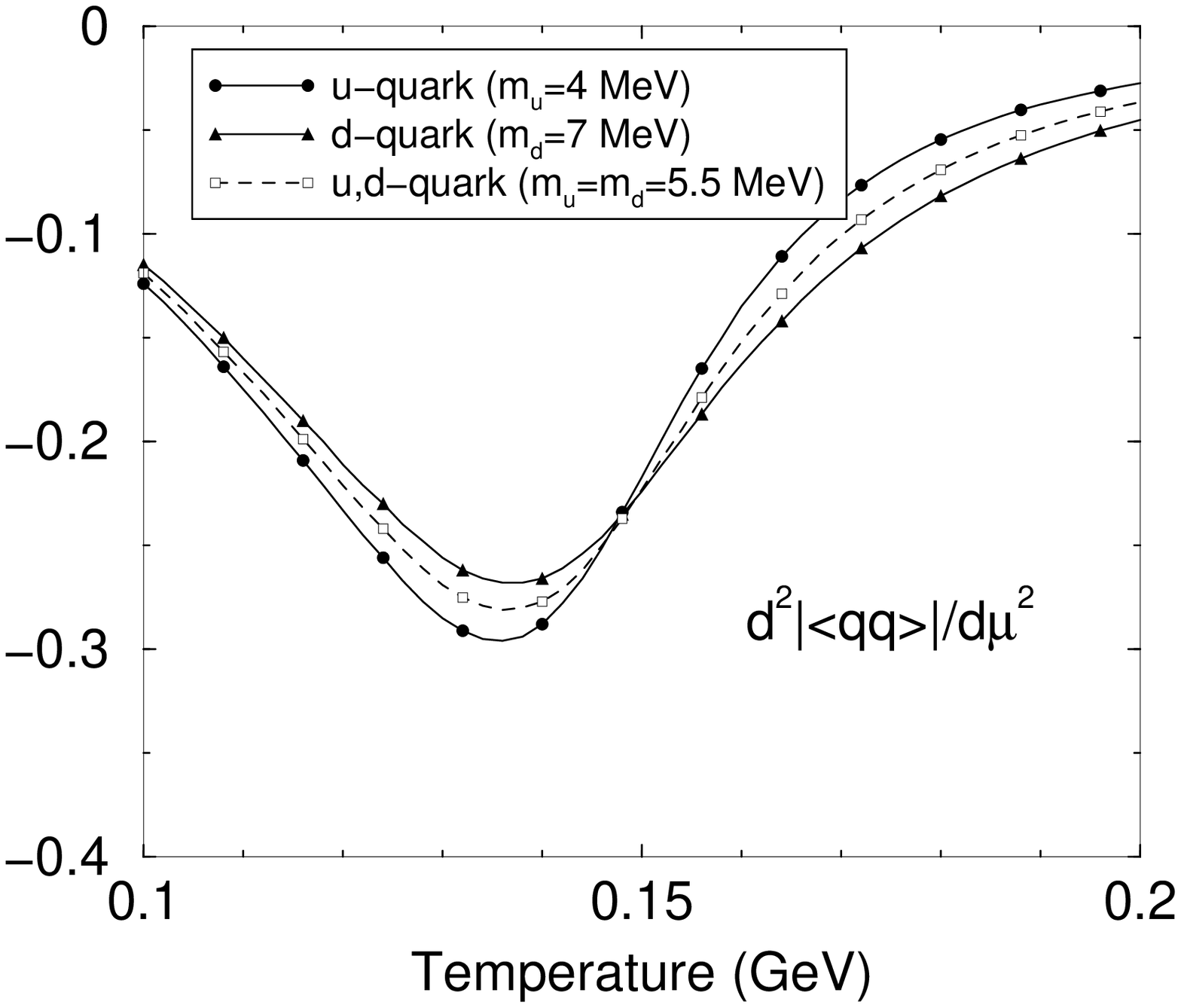}
   \vspace{0.5cm}
  \includegraphics[height=8cm,width=8cm]{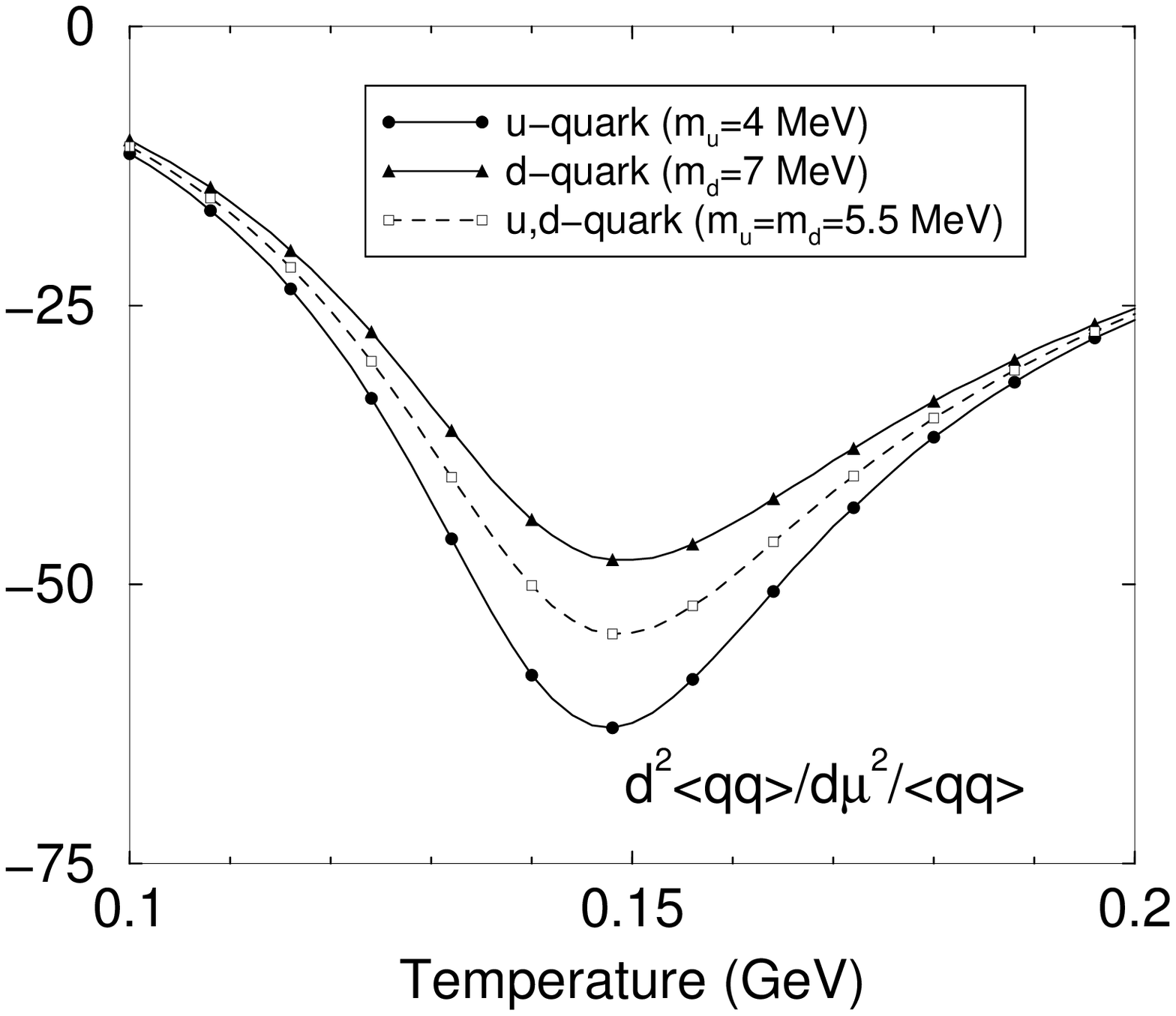}
 \caption{The second order responses of quark condensates at $\mu$ = 0:
  $\ddqcdmu$ (left) and $\ddqcdmu$/$\qc$ (right). Here,
  $\mu$ means both $\mu_S$ and $\mu_V$, i.e., $\ddqcdmus$ = $\ddqcdmuv$.}
% \vspace{1cm}
 \label{ddqc}
\end{figure*}
%--------------------
\begin{figure}[!]
 \includegraphics[height=8cm,width=8cm]{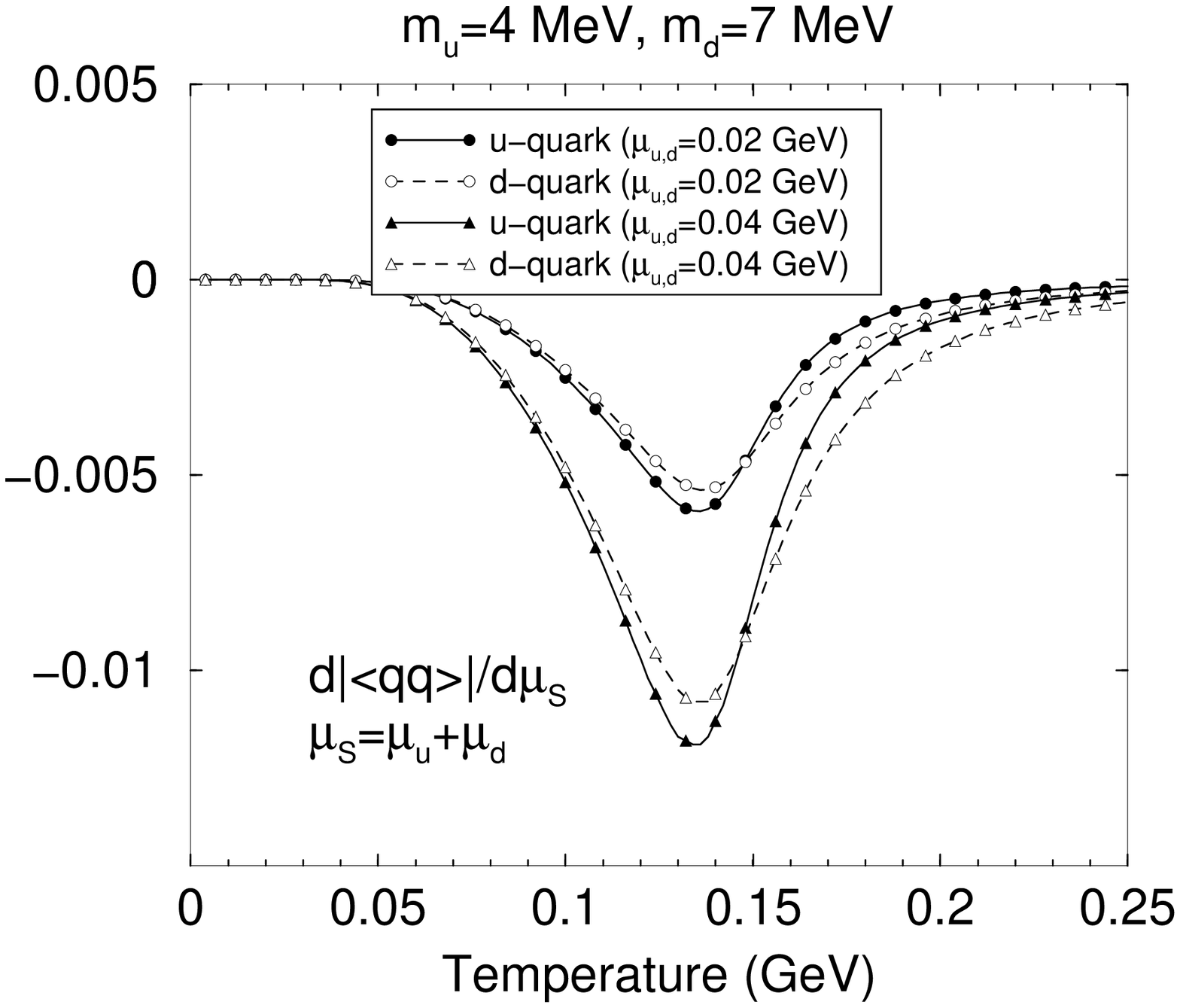}
\caption{The first order responses of the $u$ and $d$ quark
condensates at finite chemical potential, where $m_u$ = 4 MeV and
$m_d$ = 7 MeV.}
% \vspace{1cm}
 \label{dqc-ud-dm}
\end{figure}
%---------------

Let us comment on the definition of chemical potential used in the
lattice calculations and ours. There will be no difference between
the two definitions if we choose the same u- and d-quark mass. In
the NJL model calculations we chose rather general definitions for
the isoscalar and isovector chemical potentials, i.e., $\mu_S$ =
$\mu_u + \mu_d$ and $\mu_V$ = $\mu_u - \mu_d$ to show the quark
mass dependence by taking different u- and d-quark masses.

Using Eq. (\ref{dqdmu}) one can show that both $\dqcdmus$ and
$\dqcdmuv$ are zero at $\mu$=0. This is the same as the lattice
QCD calculation \cite{taro-lattice}. At finite chemical potential,
the absolute value of the quark condensate decreases with
increasing chemical potential and the variation is proportional to
the chemical potential in the present NJL model \cite{mc}.

Next, consider the second order response $\ddqcdmu$. This is
obtained by differentiating each equation in Eq. (\ref{dqdmu})
with respect to $\mu_S$ (or $\mu_V$) again. In Fig \ref{ddqc} we
show {$d^2 |\qc| \over d \mu^2$} for the $u$ and $d$ quark
condensates at $\mu$ = 0. Within the present NJL model the second
order response to the isoscalar chemical potential $\mu_S$ is the
same as that to the isovector chemical potential $\mu_V$. This is
consistent with the result in \cite{taro-lattice}. In fact, there
are a few different terms between the isoscalar and the isovector
cases in the lattice QCD simulations. However, it is found that
the contribution of those terms is negligible as shown in the
figure 3 of Ref.\cite{taro-lattice}.

In the lattice calculations the same u- and d-quark mass was taken
\cite{taro-lattice}. In this paper, however, we would like to show
the quark mass dependence of the response in the figure. Our main
interest is to compare the responses below and above the Mott
temperature depending on the quark mass. The present results may
be useful for future lattice QCD simulations which use different
u- and d-quark masses.

We take $m_u$ = 4 MeV and $m_d$ = 7 MeV as well as $m_u$ = $m_d$ =
5.5 MeV to show the quark mass dependence of {$d^2 |\qc| \over d
\mu^2$}. Although the cut-off and the coupling constants should be
modified, we use the same coupling constants and the cut-off for
both cases and study the behavior of $\ddqcdmu$. Below the pion
Mott temperature $T_{m_\pi}$ ($\approx$ 148 MeV in this work), the
response of the $u$ ($m_u$ = 4 MeV) quark condensate is larger
than that of the $d$ ($m_d$ = 7 MeV) quark condensate, while above
the Mott temperature the behavior of the response is opposite
(Here, $T_{m_\pi}$ is determined as a temperature at which the sum
of the $u$ and $d$ constituent quark masses equals to the pion
mass, i.e., $M_u + M_d = m_\pi$). This behavior results from the
quark mass dependence of the first order response at finite
chemical potential shown in Fig. \ref{dqc-ud-dm}, where we take
$\mu_u$ = $\mu_d$ = 0.02 and 0.04 GeV, respectively. That behavior
in Fig. \ref{dqc-ud-dm} is not altered even in the Case I
\cite{hk94}, where all the couplings and the cut-off are
independent of temperature and chemical potential. For example, in
the Case I and at $\mu$ = 0.02 GeV, the pion Mott temperature is
about 203 MeV and the extremum is located at around 186 MeV.

In Fig. \ref{ddqc}, we also present the relative variation
$\ddqcdmu$/$\qc$ at $\mu$=0 for comparison. Now, $\ddqcdmu$/$\qc$
has the extremum at the Mott temperature and the (absolute value
of) variation decreases with increasing the quark mass. We find
similar behavior in the lattice data, although $\qc$ is very small
over $T_c$ in this case.

As we know, we can not detect a quark condensate itself in
experiments. However, variations of quark condensates, such as
shown in Fig. \ref{dqc-ud-dm}, are necessary to predict responses
of hadron masses to the chemical potential, e.g., $\dmdmu$ in
\cite{mc}. A more detailed analysis on the extremum point and its
effects on hadron masses is in progress.

In summary, we have calculated the first and second order
responses of quark condensates to the chemical potential within
the NJL model and found that they are consistent with those from
the recent lattice QCD simulations.

%------------------------ Acknowledgements --------------------------%
\begin{acknowledgments}

 We would like to thank Prof. Su H. Lee for valuable comments. Two of the
authors (A.N. and T.T.) were supported by  the Grant-in-Aide for
Scientific Research by the Ministry of Education and Culture,
Japan (11440080, 12554008, 13135216, and 13740164), and S.C. was
supported by the Korean Ministry of Education under the BK21
program and in part by the Japan Society for the Promotion of
Science (JSPS). This paper is based on a contribution to the Prof.
Osamu Miyamura memorial symposium held at Institute for Nonlinear
Sciences and Applied Mathematics (INSAM), Hiroshima University,
Nov. 16--17, 2001.

\end{acknowledgments}
%
%------------------------ References --------------------------------%
%--------------------------------------------------------------------%

%-------------------------------------------------------------------%
\end{document}